\documentclass[12pt]{article}
\usepackage[english]{babel}
\usepackage[colorlinks,linkcolor=blue,citecolor=blue,filecolor=blue,urlcolor=blue]{hyperref}
\usepackage{graphics}
\usepackage{graphicx}
\usepackage{amsfonts}
\usepackage{amsmath}
\usepackage{bibunits}
\usepackage{xcolor}
\usepackage{multirow}
\usepackage{hhline}
\usepackage{rotating}
\usepackage{tabularx}

\usepackage{authblk}
\usepackage{cite}

\usepackage{ulem}
\usepackage{color,colortbl}

\definecolor{Gray}{gray}{0.85}

\title{Student-at-risk detection by current learning performance indicators using Bayesian networks}

\author[1]{T.~A.~Kustitskaya \thanks{E-mail: \href{mailto:m-tanika@yandex.ru}{m-tanika@yandex.ru}}}
\author[1]{A.~A.~Kytmanov \thanks{E-mail: \href{mailto:aakytm@gmail.com}{aakytm@gmail.com}}}
\author[1]{M.~V.~Noskov \thanks{E-mail: \href{mailto:mvnoskov@yandex.ru}{mvnoskov@yandex.ru}}}

\affil[1]{School of Space and Information Technology, Siberian Federal University, 79 Svobodny av., 660041 Krasnoyarsk, Russia}

\date{}

\begin{document}
\maketitle

\begin{abstract}
The present article is focused on the problem of prediction of student failures with the purpose of their possible prevention by timely introducing supportive measures. We propose a concept for building a predictive model based on Bayesian networks for an academic course or module taught in a blended learning format. Our empirical studies confirm that the proposed approach is perspective for the development of an early warning system for various stakeholders of the educational process.
\end{abstract}

\section{Introduction}


The increasing spread of mass online courses, online learning, and blended learning offers new challenges to educational institutions such as loss of awareness and control over certain parts of the learning process. At the same time, the rapid development of technology makes it possible to collect a variety of educational data which can provide one with insights into learners' behavior and ways of achieving learning outcomes by means of its thorough analysis \cite{ferguson, greller, klein}.

All of the above became prerequisites for the emergence of Learning Analytics (LA), a new branch of Data Analysis, that comprises aims and methods drawn from educational and psychological studies.

According to the most popular definition, LA is the measurement, collection, analysis, and reporting of data about learners and their contexts, for purposes of understanding and optimizing learning and the environments in which it occurs \cite{siemens2012special}.

LA is expected to provide benefits for all the stakeholders  (students, teachers, designers, administrators) of the higher education marketplace \cite{avella}. For instance, students may benefit from LA through personalized and adaptive support of their learning journey \cite{if2019}.

A considerable amount of students who continuously face demanding educational requirements and challenges of university life become unable to cope with their compulsory educational duties. This leads to numerous dropouts, especially among freshmen \cite{paura}. This problem can be addressed by various supportive measures such as introducing personalized tutoring or extra adaptation courses into the educational process. As it was mentioned in \cite{kuh}, the development of a supportive campus environment, incorporating various pedagogical approaches, validation and teaching activities might considerably improve student success rates.

However, for the effective implementation of such measures, one needs to have at their convenience a reliable tool for the early prediction of study success or, what is indeed more important, failure.

\section{Examples of studies on learning success}

There is no universal definition to ``learning (academic or study) success'' \cite{york}. One reason may be different perspectives of ``success'' by students, teaching staff, or society \cite{hommel2019}. Such things as a good final grade in a certain course, acceptable GPA, achievement of a degree, satisfaction with education, employability, development of student's professional competencies might all be considered as criteria for learning success. The relevance of student success prediction is confirmed by a large number of publications on the topic. In \cite{wong2019}, the authors state that, according to their search, over the seven years from 2011 to December 2017, 164 papers on the topic have been published in 46 journals and 33 conference proceedings indexed in the databases of Scopus or Web of Science.

In the numerous studies on the topic, one can find various approaches to the prediction of learning success. For instance, in \cite{S-V}, the authors combine logistic regression, linear discriminant analysis, and support vector machines for the success/failure prediction. They use a number of features available from the learning platform as predictors and conclude that the pace of activities (i.e., the frequency of events) performed by students in the platform used as the only predictor produces the most accurate results.

In \cite{Ornelas}, the authors focus on developing and applying the Naive Bayesian Classifier to the data from the LMS to predict the dropout rate, using such criteria of student performance as a number of inputs, time spent, weighted number of inputs, and weighted time spent as predictors.

In \cite{bekele2005}, the authors develop a Bayesian network to predict students' final grades in the course in Mathematics, using such predictors as gender, attitude to teamwork, interest in math, motivation for studying, self-confidence, shyness, the level of English.

In \cite{mac2010}, the authors develop an early warning system to identify at-risk
students, using logistic regression, which is built on such key variables as the total number of messages posted in forum discussions, the total number of email messages sent, and the total number of assignments completed.

In \cite{akccapinar}, the authors study the effectiveness of various algorithms (Naive Bayes, Classification Tree, Random Forest, Support Vector Machines, Neural Network, CN2 Rules, and k Nearest Neighbours) for the early prediction of student success using data from the online learning environment. The best classification performance was shown by the k Nearest Neighbours method and CN2 Rules. The authors also concluded that the conversion of all features into a categorical form improves the classification performance.

In \cite{Bystrova2018} and \cite{Ozerova2019}, the students are initially split into three groups (unsuccessful, successful at the minimum level, and successful students) as they undergo some initial assessment. Later on, based on their performance in the course throughout the semester, a student can be moved to another group. The authors then predict the probabilities of student transition from a certain group to another by means of the theory of Markov processes and Markov chains.

\subsection{Research on learning success at Siberian Federal University}

The educational process at the School of Space and Information Technology of Siberian Federal University (SSIT) utilizes blended learning approach based on certain principles including the one that states that constant assessment of students' results through electronic learning systems is essential for achieving their learning success (the whole set of principles, the so-called polyparadigm approach, was described in \cite{Shersh}). The in-class education is accompanied by the distant work of the students in the Moodle-based LMS ``E-Courses'' capable of collecting and storing significant amounts of data on student learning behavior and learning progress. The fact that most of the students agree to sign a data privacy statement upon entering the university, allows one to use their anonymized educational data for research purposes.

For the purpose of administrative management of the educational process in SSIT, there has been developed an automated management information system ``AIS SSIT'' consisting of several independent modules. The ``Electronic dean's office'' module, for instance, enables teaching and administrative staff to monitor student attendance, their current scores in ``E-Courses'' and final exam grades.

The data from these sources were used in some previously done studies on learning management and learning success. In \cite{Noskov2018}, the authors introduced a comprehensive student success rate, $U_X(t)$, as a function of time, which incorporates information about student total scores, attendance, and a number of ``effective'' entries into an e-course. For each student who participated in the experimental study, the rate was calculated weekly (so, $t$ is discrete) in order to apply some supportive measures to the students with low rates.

The authors proposed a predictive model for $U_X(t)$ based on the birth-death process where they used intensities of obtaining and assimilating information ($\lambda$ and $\mu$, relatively) as the parameters of the process. The authors calculated probabilities of $U_X(t_n)$ by Chapman-Kolmogorov equations for a discrete Markov chain using the information about $U_X(t_i)$ for $i<n$, obtained from AIS SSIT. The authors assumed that other formulas for calculation $U_X(t)$ different from the one they proposed could be used.

  


The approach to learning management based on the proposed formula for $U_X(t)$ looks rather beneficial for the LA stakeholders on macro-level (administrative staff) because the formula provides a simple indicator of the failure risk and helps to evaluate the overall situation.

At the same time, the existing information system does not cover all aspects of the educational process as it does not take into account the peculiarities of particular subject areas and design of courses.

A majority of the courses at the School are developed in frame of the paradigm of blended learning combining both traditional in-class studies and distant learning in ``E-courses'' LMS. While the student performance indicators obtained from the online environment are to some extent unified, assessing student in-class performance can significantly depend on a course instructor. Moreover, the corresponding data often remain unaccounted.
 
Furthermore, such unified metrics as a total score, attendance, or a number of effective entries may indicate a significantly different rate of failure risk for different disciplines. 
 
This problem can be solved by building a set of warning systems at a lower level of stakeholders (course leaders and teachers). 
 
Generally we aim at developing a concept of a flexible to course design warning system, incorporating predictive models and a complex of supportive measures providing adequate and timely support to the students whose current performance has patterns of learning failure. In this paper, we discuss one of the possible ways of building such a predictive model for learning success, which we formulate in terms of student-at-risk detection with two possible levels of success: success and failure.

\section{Bayesian approach to student-at-risk detection}\label{sec:gen_network}

One of the most important features of a predictive model along with its accuracy is its actionability \cite{gardner2018}, which means that a model should be able to react immediately to new information about student behavior. Another desired property is interpretability which provides a better understanding of the relationships between predictors and outcomes and enables to assess a level of importance of various factors for student success.


Out of a wide range of predictive models, the described above points made us opt for Bayesian networks. It is a simple and intuitive tool for describing conditional dependencies between variables, which timely takes into account new information about variables.

Bayesian network is a directed acyclic graph, whose nodes are random variables and edges represent conditional dependencies between variables.

The modeling procedure consists of the following steps:
\begin{itemize}
\item Determining a network structure
\item Specifying prior conditional distributions for the nodes
\item Evidence obtaining (getting new information on the values of the nodes)
\item Computing the posterior joint distribution for the nodes 
\end{itemize}

To make ourselves precise, we define \textit{learning success} as getting a passing grade in the final exam, and \textit{learning failure} as getting a failing grade.
We regard a student as a \textit{student-at-risk} if the probability of their passing the final exam is lower than the previously defined value $p\in(0,1)$.

Thus, the response variable for our model is binary with ``0'' value corresponding to failure and ``1'' corresponding to success, and 
the predictive model is a binary classifier, which we construct using the Bayesian network.

Although each discipline has its own set of assessment tools, which form student scores, some other non-scored indicators of learning behaviour can influence overall student performance and, thus, might provide important information for success prediction. In order to be included into the set of predictors along with the scores for learning activities, such indicators need to be converted into numerical or categorical type.

\begin{figure}[ht]
\begin{center}
\vspace{0.5cm}

\includegraphics[width=0.7\linewidth]{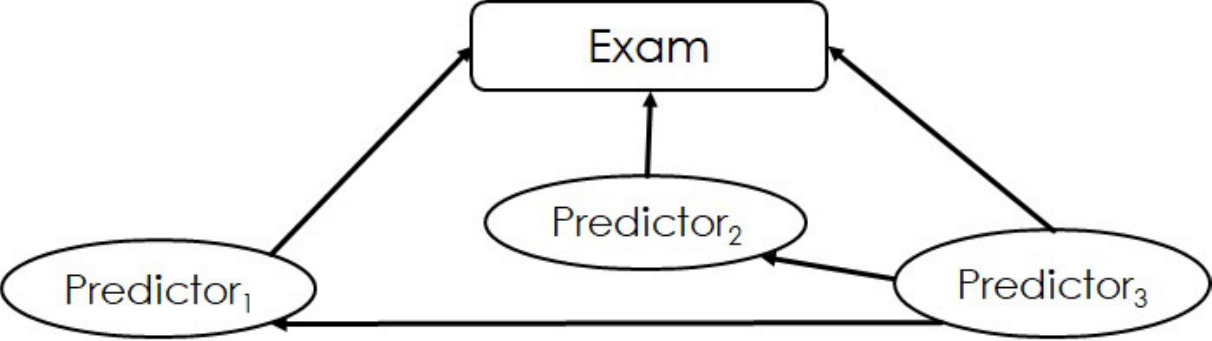}
\vspace{0.5cm}

\caption{An example of the Bayesian network structure.  
$\text{Predictor}_1$ and $\text{Predictor}_2$ are assumed to be conditionally independent;  $\text{Predictor}_1$ and $\text{Predictor}_3$, as well as $\text{Predictor}_2$ and $\text{Predictor}_3$ are conditionally dependent.}
\label{Figure:gen_structure}
\end{center}
\end{figure}

After defining predictors, we assume them and the response variable to be random variables and regard them as nodes for the Bayesian network. The known conditional dependencies between the predictors should be captured with the edges connecting the corresponding nodes. All the missing connections define conditional independencies between the variables. (See the example of the network in Figure \ref{Figure:gen_structure}). 

Prior conditional probabilities can be obtained by expert judgment or using statistical inference from a dataset which contains the data on performance of students, who have already completed the discipline 
(and hence have got the final exam grades). Having defined the prior conditional probabilities, we may regard the Bayesian network as constructed.

To predict the probability of a student success at some certain point during the semester, we need to include into the network new information on the predictor values known up to the moment. Then, we compute the posterior distribution for the response using Bayes's rule. 

If the posterior probability of success for a certain student is less than $p$, we classify them as a student-at-risk.

\section{Student-at-risk detection in the course of Probability and Statistics}

To implement the designed concept of student-at-risk detection in the educational process in the School of Space and Information Technology of Siberian Federal University and to assess its usability, a Bayesian network was developed to accompany the course of Probability and Statistics. 

Students majoring in Applied Mathematics and Information Security take this course in Spring semester. Each academic year the number of students taking the course varies from 40 to 45. The course is taught using the technology of blended learning, which means that students are obliged to attend lectures and practicums (recitations), and do out-of-class independent work in the electronic course, completing individual e-tests.

\begin{figure}[ht]
\begin{center}
\includegraphics[width=1\linewidth]{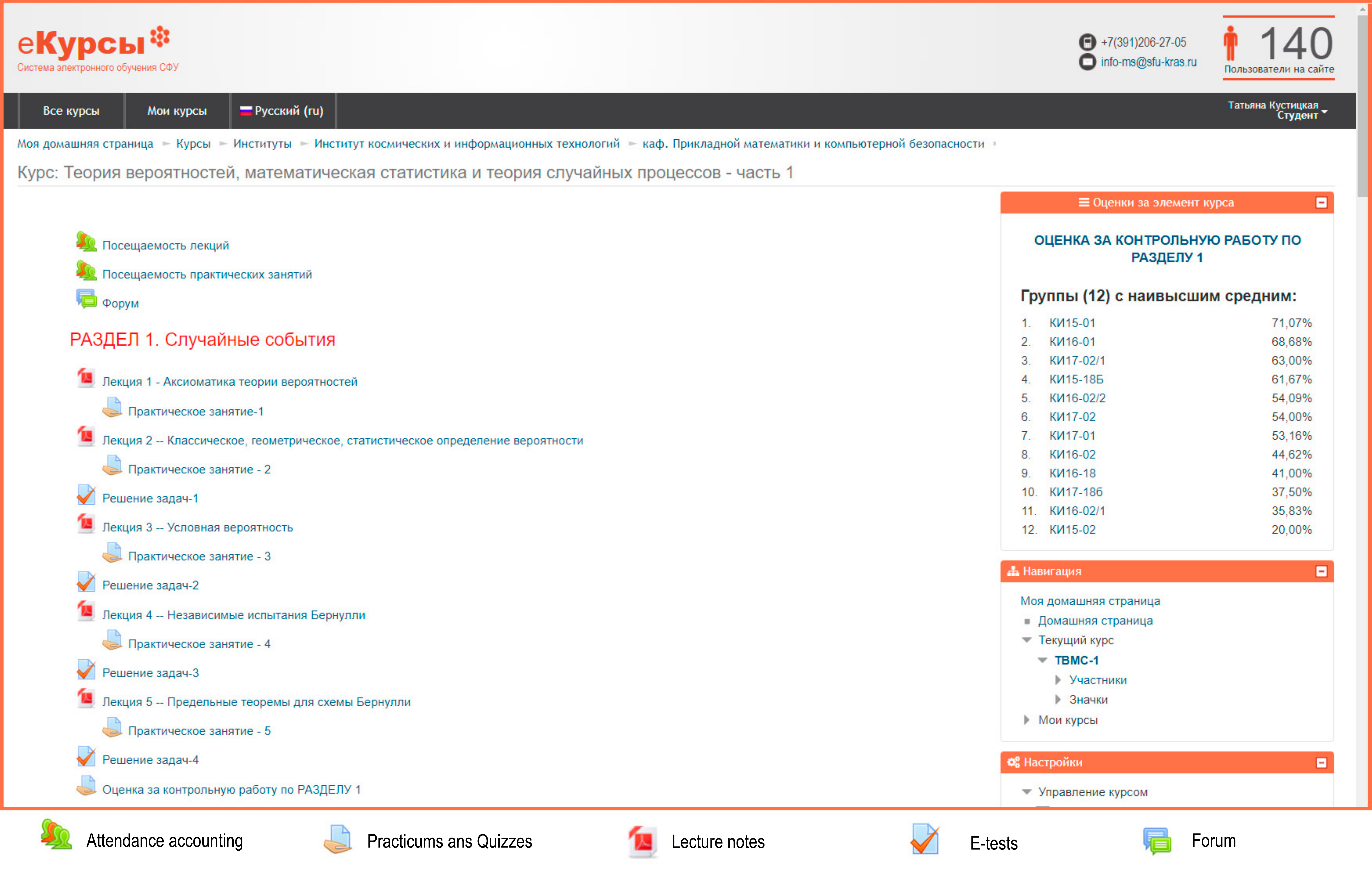}
\vspace{0.2cm}

\caption{The home page of the course Probability and Statistics}
\label{Figure:home_screen}
\end{center}
\end{figure}

Within the course students are assessed for practicums, quizzes, e-tests, attendance at lectures and practicums. All the assessment tools are integrated in the electronic course (see Figure \ref{Figure:home_screen}).  

\subsection{Preliminary student performance analysis}

The data on student performance and final grades for the course of Probability and Statistics were collected for 2016/2017,  2017/2018, and 2018/2019 academic years. The dataset was formed by the educational data for the total number of 129 students.

``E-Cources'' was the main data source for the research, providing the information on student performance in the electronic environment as well as in-class performance. The data on final grades were collected from the module ``Electronic dean's office'' of AIS SSIT.  Before the statistical analysis, all data records were anonymized and provided with new identifiers by means of a random function.

At the first stage we conducted a preliminary descriptive and visual analysis of data on students' final grades and their performance throughout the semester including attendance and grades for various types of assignments.

\begin{figure}[!ht]
\begin{center}
\includegraphics[width=0.6\linewidth]{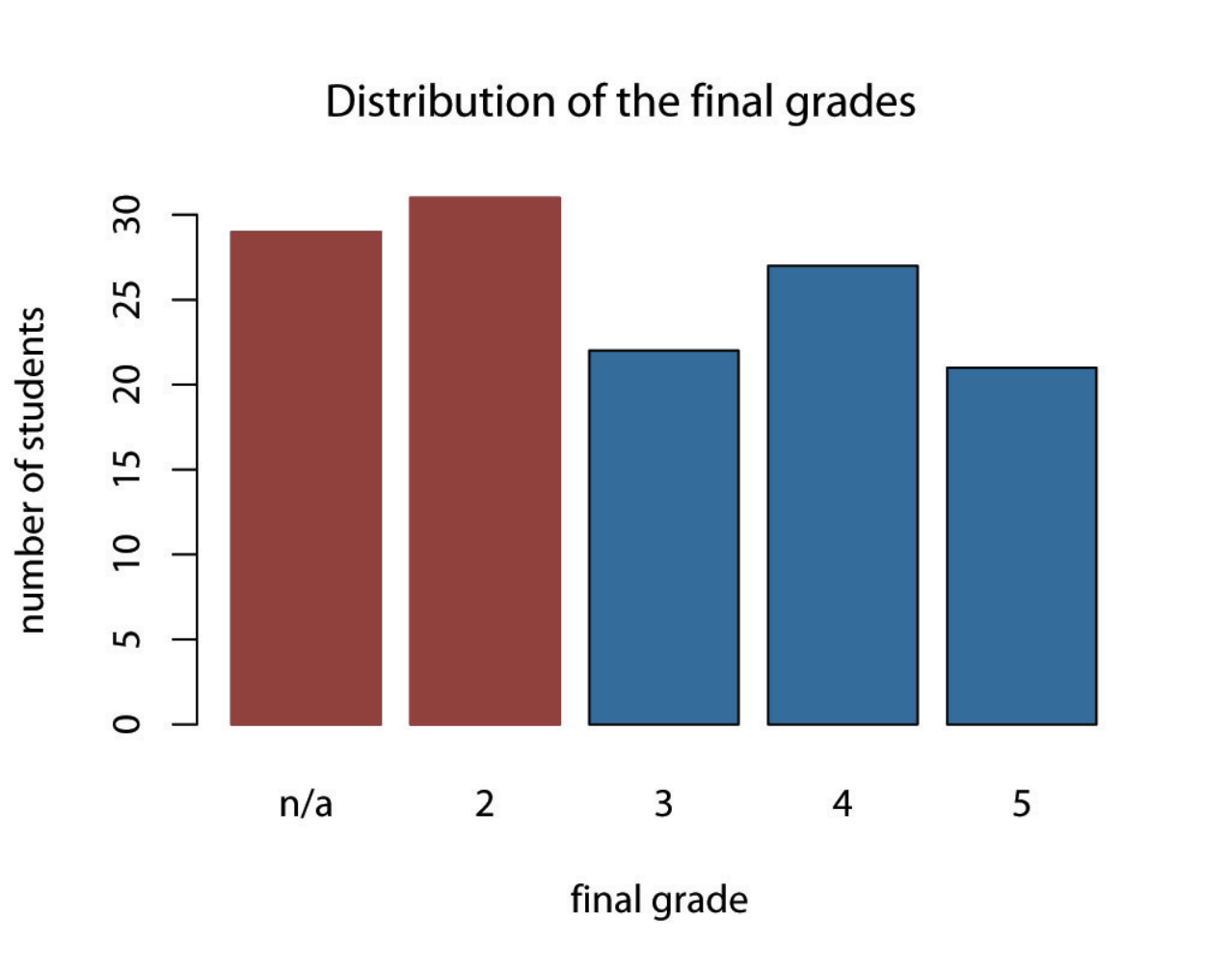}
\caption{Distribution of the final grades}
\label{Figure:exam_grades}
\end{center}
\end{figure}

There are 5 levels of grades. A student can get ``2'', ``3'', ``4'', or ``5'' for bad, satisfactory, good or excellent performance, respectively or ``n/a'' (``not awarded'') grade for not participating in a certain activity or failing to submit an assignment. The distribution of the final grades for the dataset is presented in Figure \ref{Figure:exam_grades}.

\begin{figure}[!h]
\begin{minipage}[h]{0.327\linewidth}
\center{\includegraphics[width=1\linewidth]{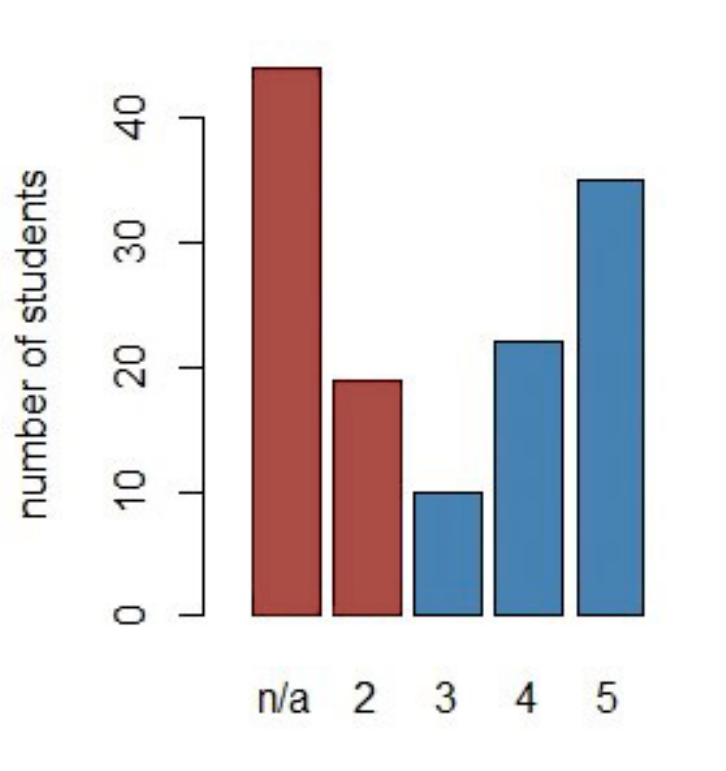} \\ Quiz 1}
\end{minipage}
\hfill
\begin{minipage}[h]{0.327\linewidth}
\center{\includegraphics[width=1\linewidth]{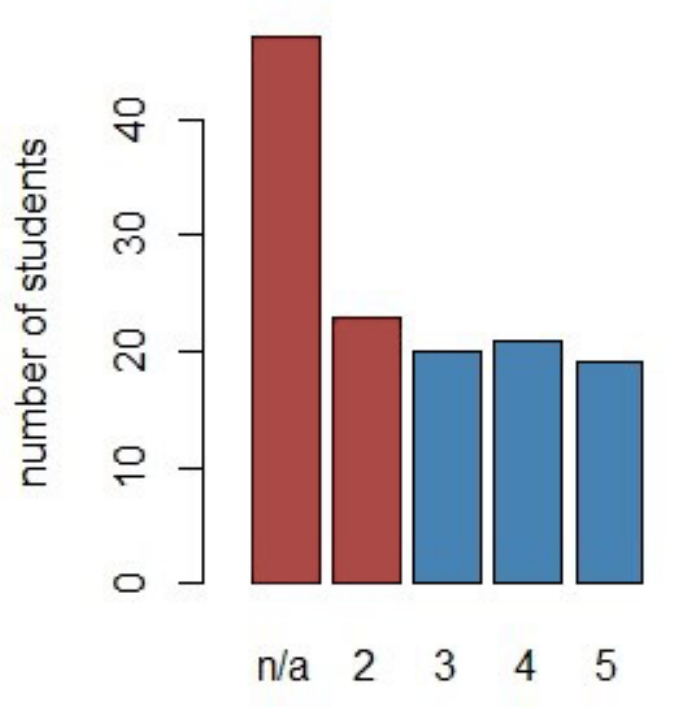} \\ Quiz 2}
\end{minipage}
\begin{minipage}[h]{0.327\linewidth}
\center{\includegraphics[width=1\linewidth]{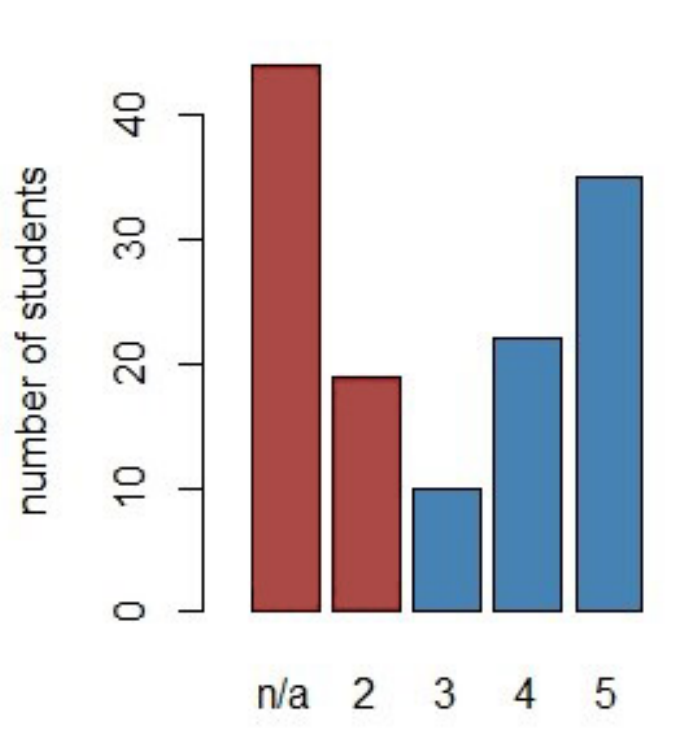} \\ Quiz 3}
\end{minipage}
\vspace{0.5cm}

\caption{Distribution of grades for quizzes}
\label{Figure:quizzes}
\end{figure}

During a semester, students take three quizzes (at Weeks 6, 13, and 17). The distribution of their grades is presented in Figure \ref{Figure:quizzes}. Grades for the quizzes are quite strongly correlated with the final grades for our dataset. The correlations between the final grades and the grades for Quizzes 1, 2, and 3 is 76\%, 79\%, and 79\% respectively. This allows us to consider the grades for the quizzes as good predictors of student success.

In addition, students take 16 electronic tests as a part of their individual work during a semester. The correlations between the e-tests scores and the final grades are pretty weak and vary from 24\% to 56\%. Together with the e-tests scores, the dataset also contains information about the number of attempts made by a student to pass a particular test. The maximum allowed number of attempts varies from 4 to 7 depending on the test. At the same time, a student who takes an e-test generally makes a small number of attempts (1 or 2). Together with test scores, numbers of attempts indicate a degree of persistence in doing individual work which can have a valuable input to success prediction.

\begin{figure}[!ht]
\begin{center}
\includegraphics[width=1\linewidth]{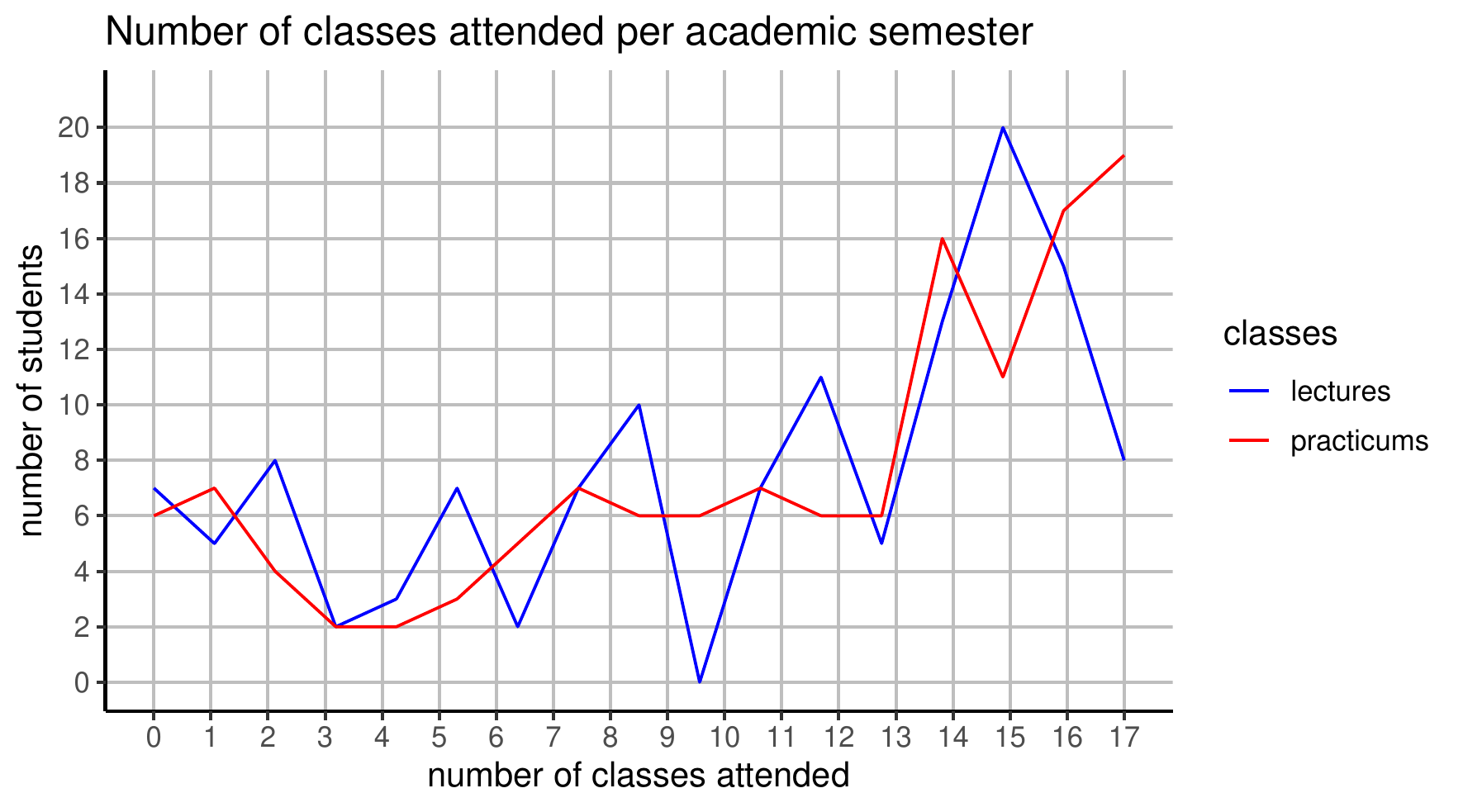}
\caption{Attendance of lectures and practicums available at the end of the course}
\label{Figure:attendance}
\end{center}
\end{figure}

Lectures and practicums are held once a week, so that we have a total number of 17 lectures and practicums. Analysis of attendance rates shows that only 61\% of students attended more than 10 lectures and 64\% attended more than 10 practicums (see Figure \ref{Figure:attendance}). Final attendance percentage (available at the end of a semester) correlates quite strongly with final grades (correlation between attendance and final grades is 69\% and 74\% for lectures and practicums, respectively). However, the intermediate correlations are considerably lower (43\% and 48\% at week 4, 59\% and 64\% at week 9), which might indicate that attendance rates are not good predictors of student success at early stages.

\subsection{Building Bayesian networks for predicting student success}

\begin{figure}[ht]
\centering
\includegraphics[width=1\linewidth]{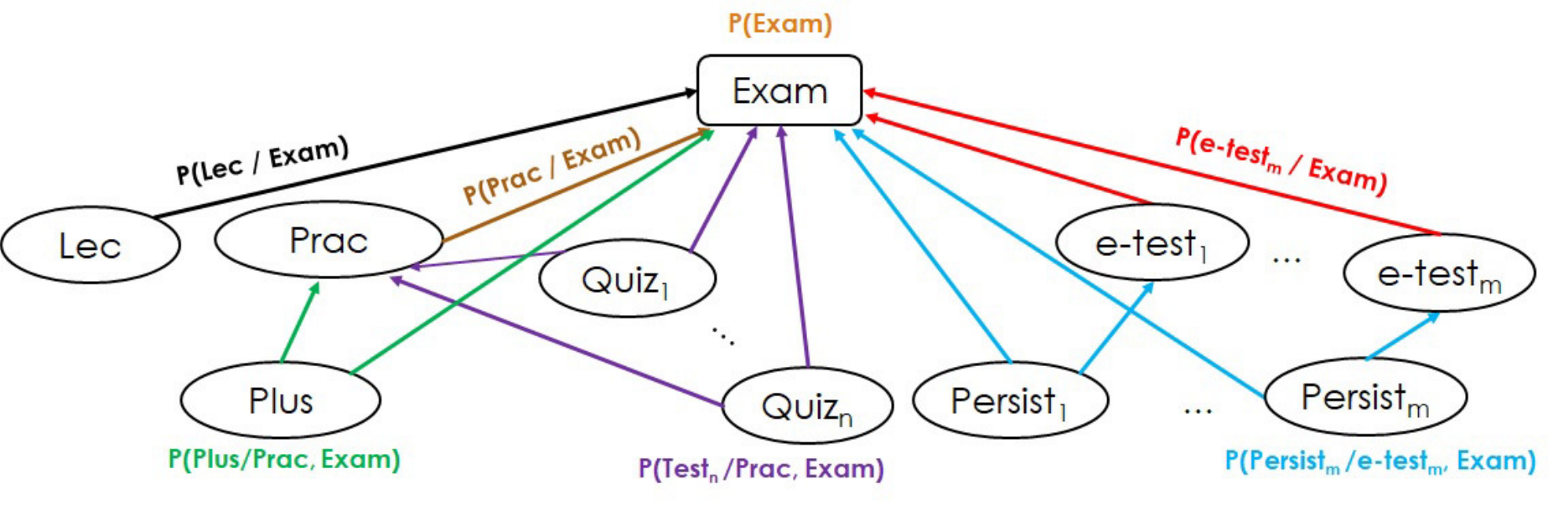}
\caption{The Bayesian network structure. }
\label{Figure:structure}
\end{figure}

We detect at-risk students during a semester by a set of Bayesian networks (Figure \ref{Figure:structure}).
As predictions are made weekly, the number of networks is equal to the number of study weeks which in our case is 17. 

Each network has the following nodes:\medskip

\textbf{A binary response variable}:
\begin{itemize}
\item \textbf{Exam} --- final exam grade (0 -- a student failed/not awarded the grade in the the exam, 1 -- a student passed the exam);
\end{itemize}

\textbf{Predictors:}

\begin{itemize}
\item \textbf{Lec} --- a number of lectures, attended by a certain week;
\item \textbf{Prac} --- a number of practicums, attended by a certain week;
\item \textbf{Plus} --- a number of points awarded for good in-class performance by a certain week;
\item $\textbf{Quiz}_i$ --- a score for the $i$-th quiz, $i=1,\ldots,n$;
\item $\textbf{e-test}_i$ --- a score for the $i$-th e-test $i=1,\dots,m$;
\item $\textbf{Persist}_i$ --- an indicator of student persistence in their work on $\text{e-test}_i$, $i=1,\ldots,m$, which depend on the numbers of attempts to pass the e-tests.
\end{itemize}

To determine each Bayesian network, we specify conditional distributions for nodes using the empirical distributions from the dataset consisting of students who have already completed the course.

At-risk students from the testing set were detected on a weekly basis using a classifier, which we call the Bayesian Network Classifier (BNC). Its operation is organized in the following way:
\begin{enumerate}
\item each week, after obtaining up-to-date data on student performance, a corresponding Bayesian network recomputes posterior distributions
\item the posterior probability of the event $\{\text{Exam}=0\}$ is compared to $p$ (a fixed cutoff of the classifier), and if it exceeds $p$, the student is classified as an at-risk student for this particular week
\end{enumerate}

\subsection{Estimation of the model efficiency in comparison with other predictive models }

To estimate a predictive efficiency of BNC in comparison with other approaches, we develop two more predictive models on the same dataset and the same set of predictors using the k-Nearest Neighbors algorithm (kNN) and the Linear Discriminant Analysis (LDA). In our case, for kNN algorithm we used $k=3$ as this number of neighbours provides the best performance of the classifier.

\begin{table}[ht]
\caption{\label{tab:conf_matrix} Confusion matrix}
\begin{center}
\begin{tabularx}{0.55\textwidth}{|>{\centering\arraybackslash}m{0.025\textwidth}|>{\centering\arraybackslash}m{0.025\textwidth}|>{\centering\arraybackslash}m{0.2\textwidth}|>{\centering\arraybackslash}m{0.2\textwidth}|}
\hhline{~~--}
\multicolumn{2}{c|}{\multirow{2}{*}{}} & \multicolumn{2}{c|}{Actual class}\\ \hhline{~~--}  \multicolumn{2}{c|}{}  &  Fails exam  &  Passes exam\\
 \hline
 \multirow{2}{*}{\begin{sideways}Predicted class \end{sideways}} & \begin{sideways}Fails exam\;\; \end{sideways} & $\begin{array}{c}\text{True Positives}\\ \text{(TP)}\end{array}$ &$\begin{array}{c} \text{False Positives}\\ \text{(FP)} \end{array}$\\\hhline{~---}
 &\begin{sideways} Passes exam\;\; \end{sideways} & $\begin{array}{c} \text{False Negatives} \\ \text{(FN)}\end{array}$&$\begin{array}{c} \text{True Negatives}\\ \text{(TN)}\end{array}$\\\hline
\end{tabularx} 
\end{center}
\end{table}


To assess whether the classifiers appear to identify successful and unsuccessful students properly, we use a number of standard classification performance metrics, which are calculated from the confusion matrix (see Table~\ref{tab:conf_matrix}), where
\begin{itemize}
\item[]\textbf{True Positives (TP)} is a number of the students correctly classified by the algorithm as those, who fails the exam.
\item[]\textbf{False Positives (FP)} is a number of the students misclassified by the algorithm as those, who fails the exam.
\item[]\textbf{False Negatives (FN)} is a number of the students misclassified by the algorithm as those, who passes the exam.
\item[]\textbf{True Negatives (TN)} is a number of the students correctly classified by the algorithm as those, who passes the exam.
\end{itemize}

To perform a thorough analysis of the quality of classification, we use the following metrics:
$$\textbf{accuracy}=\cfrac{TP+TN}{TP+TN+FP+FN},$$
$$\textbf{sensitivity}=\cfrac{TP}{TP+FN},$$
$$\textbf{precision}=\cfrac{TP}{TP+FP},$$
$$\textbf{specificity}=\cfrac{TN}{TN+FP},$$
\begin{equation}\label{weighted F-score}
\textbf{F}=\cfrac{(1+\beta^2)\cdot \text{sensitivity}\cdot \text{precision}}{(\beta^2\cdot \text{precision}+\text{sensitivity})},
\end{equation}
where $0<\beta<1$ gives more weight to precision, while $\beta>1$ gives more weight to sensitivity.\bigskip

In the current problem setting, we regard sensitivity as much more valuable measure of quality for a classifier than accuracy, precision and specificity since the most important task for the warning system is to detect all the students who are likely to fail the exam. Nevertheless, the students who perform at a good enough level should not be frequently disturbed with warning messages, and consequently the percentage of true positives should also be taken into consideration. 
We therefore set $\beta=2$ in \eqref{weighted F-score} and regard the weighted F-score as a criterion for choosing the best classifier.

It is reasonable to expect that the efficiency of the built predictive model will increase over time (i.e., as the more extensive evidence about student performance the model gets). At the same time, the earlier we can detect at-risk students, the more effectively we can bring supportive measures in their study process. Thus, among the chosen models of classification we prefer the one, whose sensitivity and weighted F-score reach acceptable values at earlier stages.

To compare performance of the classifiers, we form fifteen testing sets by randomly mixing the data from the original dataset. On each set we train BNC, kNN and LDA models and estimate the quality of classification using a cross-validation procedure.

In Tables \ref{tab:pres} -- \ref{tab:F}, one can see the dynamics over time of the classifier performance metrics calculated by cross-validation on four (randomly taken) out of the fifteen testing sets.

We start with less important metrics. Comparing values of precision, specificity and accuracy for BNC, kNN and LDA models, we see (Tables~\ref{tab:pres}, \ref{tab:spec}, \ref{tab:acc}) that for the four chosen testing tests the best results for the most cases are provided by kNN algorithm. 

However, for the metrics, which we regard as more important for our problem statement, BNC shows the best results. Indeed, in Table \ref{tab:sens}, the sensitivity of BNC exceeds the sensitivity of the other algorithms in the vast majority of cases starting from the very beginning of the semester, and takes values in the interval \([0.73, 1]\). A similar picture one can observe in Table~\ref{tab:F} where BNC algorithm provides the best classification quality in terms of the weighted F-measure starting from the fifth week.

\begin{table}[htb]
\caption{\label{tab:pres} Precision for BNC, kNN and LDA on 4 out of the 15 testing sets. Maximum values of the metrics on each testing set are highlighted in gray}
\centering
\begin{tabular}[t]{|c||r|r|r||r|r|r||r|r|r||r|r|r||}
\hline
& \multicolumn{3}{c||}{set1} & \multicolumn{3}{c||}{set 2} & \multicolumn{3}{c||}{set 3} & \multicolumn{3}{c||}{set 4}\\
\cline{2-13}
\raisebox{1.5ex}[0cm][0cm] {week}
&
   BNC & kNN & LDA & BNC & kNN & LDA & BNC & kNN & LDA & BNC & kNN & LDA\\
 
\hline
   1 & 0.49 &\cellcolor{Gray} 0.94 & 0.78 & 0.50 & \cellcolor{Gray}0.97 & 0.88 & 0.49 &\cellcolor{Gray} 0.91 & 0.80 & 0.49 &\cellcolor{Gray} 1.00 & 0.78\\
\hline
   2 & 0.49 &\cellcolor{Gray} 0.91 & 0.76 & 0.50 & \cellcolor{Gray}0.92 & 0.78 & 0.49 & \cellcolor{Gray}0.89 & 0.77 & 0.47 &\cellcolor{Gray} 0.91 & 0.77\\
\hline
   3 & 0.61 & \cellcolor{Gray}0.90 & 0.80 & 0.63 & \cellcolor{Gray}0.88 & 0.76 & 0.69 &\cellcolor{Gray} 0.86 & 0.84 & 0.62 &\cellcolor{Gray} 0.94 & 0.79\\
\hline
   4 & 0.78 & \cellcolor{Gray}0.86 & 0.77 & 0.66 & \cellcolor{Gray}0.86 & 0.81 & 0.73 & \cellcolor{Gray}0.88 & 0.77 & 0.76 & \cellcolor{Gray}0.85 & 0.82\\
\hline
   5 & 0.75 & \cellcolor{Gray}0.89 & 0.81 & 0.83 & \cellcolor{Gray}0.92 & 0.77 & 0.82 & 0.85 &\cellcolor{Gray} 0.86 & 0.77 &\cellcolor{Gray} 0.88 & 0.75\\
\hline
   6 & 0.81 & \cellcolor{Gray}0.88 & 0.82 & 0.84 & \cellcolor{Gray}0.91 & 0.80 & 0.82 &\cellcolor{Gray} 0.88 & 0.86 & 0.74 &\cellcolor{Gray} 0.90 & 0.76\\
\hline
   7 & 0.74 & \cellcolor{Gray}0.88 & 0.84 & 0.81 & \cellcolor{Gray}0.88 & 0.83 & 0.77 & \cellcolor{Gray}0.88 & 0.86 & 0.72 &\cellcolor{Gray} 0.90 & 0.81\\
\hline
   8 & 0.77 & \cellcolor{Gray}0.89 & 0.87 & 0.83 & \cellcolor{Gray}0.88 & 0.83 & 0.86 &\cellcolor{Gray} 0.92 & 0.89 & 0.69 & \cellcolor{Gray}0.92 & 0.88\\
\hline
   9 & 0.79 & \cellcolor{Gray}0.92 & 0.85 & 0.89 & 0.89 & \cellcolor{Gray}0.90 & 0.86 & \cellcolor{Gray}0.94 & 0.92 & 0.72 & \cellcolor{Gray}0.92 & 0.90\\
\hline
   10 & 0.82 & \cellcolor{Gray}0.89 & 0.80 &\cellcolor{Gray} 0.88 &\cellcolor{Gray} 0.88 & 0.84 & 0.85 &\cellcolor{Gray} 0.94 & 0.85 & 0.73 & \cellcolor{Gray}0.92 & 0.89\\
\hline
   11 & 0.82 & \cellcolor{Gray}0.94 & 0.80 &\cellcolor{Gray} 0.91 & 0.87 & 0.85 & 0.81 &\cellcolor{Gray} 0.94 & 0.87 & 0.80 &\cellcolor{Gray} 0.93 & 0.91\\
\hline
   12 & 0.85 & \cellcolor{Gray}0.96 & 0.80 &\cellcolor{Gray} 0.92 & 0.86 & 0.85 & 0.81 &\cellcolor{Gray} 0.91 & 0.86 & 0.76 &\cellcolor{Gray} 0.93 & 0.92\\
\hline
   13 & 0.83 & \cellcolor{Gray}0.95 & 0.80 & \cellcolor{Gray}0.91 & 0.87 & 0.79 & 0.80 &\cellcolor{Gray} 0.94 & 0.86 & 0.79 & \cellcolor{Gray}0.93 & 0.89\\
\hline
   14 & 0.89 & \cellcolor{Gray}0.93 & 0.80 &\cellcolor{Gray} 0.95 & 0.87 & 0.84 & 0.82 &\cellcolor{Gray} 0.94 & 0.87 & 0.79 &\cellcolor{Gray} 0.93 & 0.89\\
\hline
   15 & 0.84 & \cellcolor{Gray}0.96 & 0.81 & 0.84 & \cellcolor{Gray}0.90 & 0.84 & 0.72 &\cellcolor{Gray} 0.94 & 0.88 & 0.75 &\cellcolor{Gray} 0.94 & 0.85\\
\hline
   16 & 0.88 & \cellcolor{Gray}0.92 & 0.82 &\cellcolor{Gray} 0.91 & 0.88 & 0.84 & 0.78 &\cellcolor{Gray} 0.94 & 0.88 & 0.72 & \cellcolor{Gray}0.94 & 0.83\\
\hline
   17 & 0.81 & \cellcolor{Gray}0.93 & 0.81 &\cellcolor{Gray} 0.92 & 0.91 & 0.82 & 0.79 &\cellcolor{Gray} 0.91 & 0.87 & 0.81 &\cellcolor{Gray} 0.94 & 0.85\\
\hline
\end{tabular}
\end{table}

\begin{table}
\caption{\label{tab:spec} Specificity for BNC, kNN and LDA on 4 out of the 15 testing sets. Maximum values of the metrics on each testing set are highlighted in gray}
\centering
\begin{tabular}[t]{|c||r|r|r||r|r|r||r|r|r||r|r|r||}
\hline
& \multicolumn{3}{c||}{set1} & \multicolumn{3}{c||}{set 2} & \multicolumn{3}{c||}{set 3} & \multicolumn{3}{c||}{set 4}\\
\cline{2-13}
\raisebox{1.5ex}[0cm][0cm] {week}
&
   BNC & kNN & LDA & BNC & kNN & LDA & BNC & kNN & LDA & BNC & kNN & LDA\\
 
\hline
    1 & \cellcolor{Gray}1.00 & 0.90 & 0.57 & 0.83 &\cellcolor{Gray} 0.94 & 0.31 &\cellcolor{Gray} 1.00 & 0.92 & 0.49 &\cellcolor{Gray} 1.00 &\cellcolor{Gray} 1.00 & 0.44\\
\hline
    2 & \cellcolor{Gray}1.00 & 0.90 & 0.62 & 0.84 &\cellcolor{Gray} 0.90 & 0.49 & \cellcolor{Gray}1.00 & 0.86 & 0.55 &\cellcolor{Gray} 1.00 & 0.81 & 0.68\\
\hline
    3 & 0.59 & \cellcolor{Gray}0.84 & 0.73 & 0.61 & \cellcolor{Gray}0.87 & 0.62 & 0.64 &\cellcolor{Gray} 0.79 & 0.78 & 0.63 & \cellcolor{Gray}0.83 & 0.70\\
\hline
    4 & 0.72 & \cellcolor{Gray}0.76 & 0.65 & 0.60 &\cellcolor{Gray} 0.70 & 0.58 & 0.70 & \cellcolor{Gray}0.83 & 0.63 & 0.72 &\cellcolor{Gray} 0.82 & 0.71\\
\hline
    5 & 0.75 & \cellcolor{Gray}0.83 & 0.71 & 0.76 &\cellcolor{Gray} 0.87 & 0.58 & 0.76 &\cellcolor{Gray} 0.78 & 0.77 & 0.75 &\cellcolor{Gray} 0.85 & 0.57\\
\hline
    6 & 0.79 & \cellcolor{Gray}0.82 & 0.71 & 0.82 & \cellcolor{Gray}0.86 & 0.60 &\cellcolor{Gray} 0.85 & 0.84 & 0.78 & 0.76 & \cellcolor{Gray}0.88 & 0.62\\
\hline
    7 & 0.74 &\cellcolor{Gray} 0.81 & 0.76 & \cellcolor{Gray}0.74 &\cellcolor{Gray} 0.74 & 0.61 & 0.74 & 0.74 &\cellcolor{Gray} 0.78 & 0.72 & \cellcolor{Gray}0.85 & 0.71\\
\hline
    8 & 0.76 & 0.78 &\cellcolor{Gray} 0.79 & \cellcolor{Gray}0.78 & 0.76 & 0.74 & 0.82 & \cellcolor{Gray}0.86 & 0.84 & 0.68 & \cellcolor{Gray}0.87 & 0.85\\
\hline
    9 & 0.78 & \cellcolor{Gray}0.87 & 0.74 & 0.83 &\cellcolor{Gray} 0.85 & 0.83 & 0.84 & \cellcolor{Gray}0.89 & \cellcolor{Gray}0.89 & 0.78 &\cellcolor{Gray} 0.88 & 0.86\\
\hline
    10 & 0.78 &\cellcolor{Gray} 0.79 & 0.76 &\cellcolor{Gray} 0.83 & 0.78 & 0.80 & 0.84 &\cellcolor{Gray} 0.89 & 0.74 & 0.70 & \cellcolor{Gray}0.88 & 0.82\\
\hline
    11 & 0.83 & \cellcolor{Gray}0.85 & 0.72 & \cellcolor{Gray}0.86 & 0.79 & 0.78 & 0.85 & \cellcolor{Gray}0.89 & 0.77 & 0.79 & \cellcolor{Gray}0.88 & 0.85\\
\hline
    12 & 0.86 &\cellcolor{Gray} 0.92 & 0.75 & \cellcolor{Gray}0.88 & 0.77 & 0.75 & \cellcolor{Gray}0.86 & 0.83 & 0.78 & 0.78 & \cellcolor{Gray}0.88 & 0.86\\
\hline
    13 & \cellcolor{Gray}0.88 &\cellcolor{Gray} 0.88 & 0.75 &\cellcolor{Gray} 0.88 & 0.80 & 0.72 & 0.85 &\cellcolor{Gray} 0.90 & 0.77 & 0.79 & \cellcolor{Gray}0.89 & 0.82\\
\hline
    14 & \cellcolor{Gray}0.89 &\cellcolor{Gray} 0.89 & 0.78 &\cellcolor{Gray} 0.90 & 0.80 & 0.76 & 0.88 &\cellcolor{Gray} 0.90 & 0.80 & 0.81 &\cellcolor{Gray} 0.88 &\cellcolor{Gray} 0.88\\
\hline
    15 & 0.80 &\cellcolor{Gray} 0.92 & 0.77 &\cellcolor{Gray} 0.84 & 0.83 & 0.75 & 0.74 & \cellcolor{Gray}0.90 & 0.80 & 0.73 & \cellcolor{Gray}0.91 & 0.84\\
\hline
    16 & \cellcolor{Gray}0.88 & 0.87 & 0.76 &\cellcolor{Gray} 0.86 & 0.78 & 0.75 & 0.83 & \cellcolor{Gray}0.90 & 0.79 & 0.75 & \cellcolor{Gray}0.90 & 0.80\\
\hline
    17 & 0.81 & \cellcolor{Gray}0.88 & 0.74 &\cellcolor{Gray} 0.89 & 0.84 & 0.71 & 0.79 & \cellcolor{Gray}0.82 & 0.79 & 0.77 & \cellcolor{Gray}0.90 & 0.81\\
\hline
\end{tabular}
\end{table}

\begin{table}
\caption{\label{tab:acc} Accuracy for BNC, kNN and LDA on 4 out of the 15 testing sets. Maximum values of the metrics on each testing set are highlighted in gray}
\centering
\begin{tabular}[t]{|c||r|r|r||r|r|r||r|r|r||r|r|r||}
\hline
& \multicolumn{3}{c||}{set1} & \multicolumn{3}{c||}{set 2} & \multicolumn{3}{c||}{set 3} & \multicolumn{3}{c||}{set 4}\\
\cline{2-13}
\raisebox{1.5ex}[0cm][0cm] {week}
&
   BNC & kNN & LDA & BNC & kNN & LDA & BNC & kNN & LDA & BNC & kNN & LDA\\
 
\hline
    1 & 0.42 &\cellcolor{Gray} 0.96 & 0.60 & 0.46 & \cellcolor{Gray}0.93 & 0.58 & 0.42 & \cellcolor{Gray}0.96 & 0.58 & 0.42 &\cellcolor{Gray} 0.99 & 0.58\\
\hline
    2 & 0.42 &\cellcolor{Gray} 0.91 & 0.66 & 0.46 &\cellcolor{Gray} 0.93 & 0.57 & 0.42 &\cellcolor{Gray} 0.93 & 0.66 & 0.40 & \cellcolor{Gray}0.94 & 0.65\\
\hline
    3 & 0.68 & \cellcolor{Gray}0.87 & 0.74 & 0.62 &\cellcolor{Gray} 0.87 & 0.69 & 0.67 & \cellcolor{Gray}0.84 & 0.77 & 0.63 & \cellcolor{Gray}0.87 & 0.74\\
\hline
    4 & 0.72 & \cellcolor{Gray}0.77 & 0.66 & 0.61 &\cellcolor{Gray} 0.75 & 0.68 & 0.65 & \cellcolor{Gray}0.79 & 0.71 & 0.64 & \cellcolor{Gray}0.77 & 0.67\\
\hline
    5 & 0.71 & \cellcolor{Gray}0.82 & 0.71 & 0.70 & \cellcolor{Gray}0.81 & 0.63 & 0.70 & \cellcolor{Gray}0.75 & 0.71 & 0.69 &\cellcolor{Gray} 0.74 & 0.66\\
\hline
    6 & 0.73 & \cellcolor{Gray}0.80 & 0.71 & 0.73 &\cellcolor{Gray} 0.79 & 0.66 & 0.72 &\cellcolor{Gray} 0.79 & 0.71 & 0.67 & \cellcolor{Gray}0.77 & 0.67\\
\hline
    7 & 0.69 & \cellcolor{Gray}0.79 & 0.75 & 0.68 & \cellcolor{Gray}0.73 & 0.68 & 0.71 &\cellcolor{Gray} 0.75 & 0.74 & 0.64 &\cellcolor{Gray} 0.75 & 0.72\\
\hline
    8 & 0.68 & 0.80 &\cellcolor{Gray} 0.81 & 0.71 &\cellcolor{Gray} 0.74 &\cellcolor{Gray} 0.74 & 0.73 & \cellcolor{Gray}0.79 &\cellcolor{Gray} 0.79 & 0.66 & 0.78 & \cellcolor{Gray}0.80\\
\hline
    9 & 0.70 & \cellcolor{Gray}0.81 & 0.79 & 0.74 & 0.78 &\cellcolor{Gray} 0.80 & 0.76 &\cellcolor{Gray} 0.83 & 0.81 & 0.71 & 0.81 & \cellcolor{Gray}0.83\\
\hline
    10 & 0.78 & 0.79 & \cellcolor{Gray}0.81 & 0.77 & 0.75 &\cellcolor{Gray} 0.79 & 0.82 &\cellcolor{Gray} 0.84 & 0.80 & 0.75 & 0.80 & \cellcolor{Gray}0.81\\
\hline
    11 & 0.77 &\cellcolor{Gray} 0.82 & 0.80 &\cellcolor{Gray} 0.79 & 0.75 & 0.78 & 0.79 & \cellcolor{Gray}0.84 &\cellcolor{Gray} 0.84 &\cellcolor{Gray} 0.82 & 0.81 &\cellcolor{Gray} 0.82\\
\hline
    12 & 0.79 &\cellcolor{Gray} 0.83 & 0.81 &\cellcolor{Gray} 0.81 & 0.78 & 0.79 & 0.81 &\cellcolor{Gray} 0.83 & 0.81 & 0.78 & \cellcolor{Gray}0.82 & 0.81\\
\hline
    13 & 0.81 & \cellcolor{Gray}0.84 & 0.81 &\cellcolor{Gray} 0.81 & 0.76 & 0.78 & 0.81 &\cellcolor{Gray} 0.84 & 0.78 & 0.81 & \cellcolor{Gray}0.83 & 0.81\\
\hline
    14 & 0.81 & \cellcolor{Gray}0.84 & 0.83 & \cellcolor{Gray}0.83 & 0.76 & 0.83 & 0.83 &\cellcolor{Gray} 0.87 & 0.85 & 0.81 & 0.84 & \cellcolor{Gray}0.85\\
\hline
    15 & 0.79 & \cellcolor{Gray}0.86 & 0.83 & 0.72 & 0.78 &\cellcolor{Gray} 0.80 & 0.75 & \cellcolor{Gray}0.87 & 0.85 & 0.76 & \cellcolor{Gray}0.84 & 0.83\\
\hline
    16 & 0.81 & \cellcolor{Gray}0.84 & 0.83 &\cellcolor{Gray} 0.79 & 0.77 &\cellcolor{Gray} 0.79 & 0.81 & \cellcolor{Gray}0.87 & 0.81 & 0.74 &\cellcolor{Gray} 0.83 & 0.81\\
\hline
    17 & 0.75 & \cellcolor{Gray}0.85 & 0.81 & 0.76 &\cellcolor{Gray} 0.80 & 0.77 & 0.78 & \cellcolor{Gray}0.84 & 0.80 & 0.74 &\cellcolor{Gray} 0.85 & 0.82\\
\hline
\end{tabular}
\end{table}



\begin{table}
\caption{\label{tab:sens} Sensitivity for BNC, kNN and LDA on 4 out of the 15 testing sets. Maximum values of the metrics on each testing set are highlighted in gray}
\centering
\begin{tabular}[t]{|c||r|r|r||r|r|r||r|r|r||r|r|r||}
\hline
& \multicolumn{3}{c||}{set1} & \multicolumn{3}{c||}{set 2} & \multicolumn{3}{c||}{set 3} & \multicolumn{3}{c||}{set 4}\\
\cline{2-13}
\raisebox{1.5ex}[0cm][0cm] {week}
&
   BNC & kNN & LDA & BNC & kNN & LDA & BNC & kNN & LDA & BNC & kNN & LDA\\
 
\hline
 1 & \cellcolor{Gray}1.00 &\cellcolor{Gray} 1.00 & 0.38 & 0.93 &\cellcolor{Gray} 0.94 & 0.16 & \cellcolor{Gray}1.00 &\cellcolor{Gray} 1.00 & 0.22 &\cellcolor{Gray} 1.00 & 0.98 & 0.27\\
\hline
 2 & \cellcolor{Gray}1.00 & 0.91 & 0.47 & \cellcolor{Gray}1.00 & 0.97 & 0.31 &\cellcolor{Gray} 1.00 & 0.98 & 0.42 & \cellcolor{Gray}0.97 & 0.95 & 0.47\\
\hline
 3 & \cellcolor{Gray}0.88 & 0.85 & 0.70 & 0.80 & \cellcolor{Gray}0.86 & 0.59 &\cellcolor{Gray} 0.94 & 0.85 & 0.65 & \cellcolor{Gray}0.83 & 0.76 & 0.68\\
\hline
 4 & \cellcolor{Gray}0.85 & 0.69 & 0.58 & \cellcolor{Gray}0.73 & 0.67 & 0.53 &\cellcolor{Gray} 0.83 & 0.68 & 0.53 &\cellcolor{Gray} 0.85 & 0.58 & 0.46\\
\hline
 5 & \cellcolor{Gray}0.77 & 0.76 & 0.67 & \cellcolor{Gray}0.84 & 0.68 & 0.46 &\cellcolor{Gray} 0.88 & 0.60 & 0.46 &\cellcolor{Gray} 0.90 & 0.48 & 0.50\\
\hline
 6 & \cellcolor{Gray}0.83 & 0.74 & 0.64 & \cellcolor{Gray}0.86 & 0.66 & 0.51 &\cellcolor{Gray} 0.88 & 0.68 & 0.52 &\cellcolor{Gray} 0.83 & 0.54 & 0.50\\
\hline
 7 & \cellcolor{Gray}0.75 & 0.66 & 0.66 & \cellcolor{Gray}0.79 & 0.54 & 0.48 &\cellcolor{Gray} 0.83 & 0.57 & 0.63 &\cellcolor{Gray} 0.80 & 0.49 & 0.56\\
\hline
 8 & \cellcolor{Gray}0.78 & 0.69 & 0.72 & \cellcolor{Gray}0.89 & 0.57 & 0.66 &\cellcolor{Gray} 0.89 & 0.65 & 0.78 & \cellcolor{Gray}0.74 & 0.54 & 0.67\\
\hline
 9 & \cellcolor{Gray}0.78 & 0.69 & 0.71 & \cellcolor{Gray}0.87 & 0.62 & 0.67 &\cellcolor{Gray} 0.90 & 0.66 & 0.72 &\cellcolor{Gray} 0.76 & 0.64 & 0.69\\
\hline
 10 & \cellcolor{Gray}0.88 & 0.69 & 0.82 & \cellcolor{Gray}0.89 & 0.61 & 0.73 &\cellcolor{Gray} 0.98 & 0.69 & 0.73 &\cellcolor{Gray} 0.86 & 0.61 & 0.70\\
\hline
 11 & \cellcolor{Gray}0.85 & 0.66 & 0.78 & \cellcolor{Gray}0.86 & 0.59 & 0.73 &\cellcolor{Gray} 0.90 & 0.70 & 0.79 &\cellcolor{Gray} 0.90 & 0.64 & 0.72\\
\hline
 12 & \cellcolor{Gray}0.85 & 0.66 & 0.82 & \cellcolor{Gray}0.86 & 0.68 & 0.76 &\cellcolor{Gray} 0.90 & 0.72 & 0.80 & \cellcolor{Gray}0.85 & 0.66 & 0.63\\
\hline
 13 & \cellcolor{Gray}0.85 & 0.70 & 0.82 & \cellcolor{Gray}0.85 & 0.64 & 0.77 &\cellcolor{Gray} 0.92 & 0.70 & 0.77 &\cellcolor{Gray} 0.88 & 0.69 & 0.68\\
\hline
 14 & 0.85 & 0.70 &\cellcolor{Gray} 0.87 & \cellcolor{Gray}0.89 & 0.64 & 0.83 &\cellcolor{Gray} 0.84 & 0.74 & 0.75 &\cellcolor{Gray} 0.86 & 0.75 & 0.79\\
\hline
 15 & \cellcolor{Gray}0.91 & 0.72 & 0.87 & \cellcolor{Gray}0.78 & 0.64 & 0.76 &\cellcolor{Gray} 0.80 & 0.74 & 0.78 & \cellcolor{Gray}0.83 & 0.72 & 0.82\\
\hline
 16 & \cellcolor{Gray}0.87 & 0.72 & 0.87 & \cellcolor{Gray}0.86 & 0.64 & 0.75 &\cellcolor{Gray} 0.85 & 0.74 & 0.75 &\cellcolor{Gray} 0.84 & 0.68 & 0.79\\
\hline
 17 & \cellcolor{Gray}0.85 & 0.76 & 0.81 & \cellcolor{Gray}0.78 & 0.68 & 0.70 &\cellcolor{Gray} 0.83 & 0.74 & 0.71 &\cellcolor{Gray} 0.89 & 0.74 & 0.78\\
\hline
\end{tabular}
\end{table}

\begin{table}

\caption{\label{tab:F}Weighted F-score for BNC, kNN and LDA on 4 out of 15 testing sets. Maximum values of the metrics on each testing set are highlighted in gray}
\centering
\begin{tabular}[t]{|p{0.75cm}||r|r|r||r|r|r||r|r|r||r|r|r||}
\hline
& \multicolumn{3}{c||}{set1} & \multicolumn{3}{c||}{set 2} & \multicolumn{3}{c||}{set 3} & \multicolumn{3}{c||}{set 4}\\
\cline{2-13}
\raisebox{1.5ex}[0cm][0cm] {week}
&
   BNC & kNN & LDA & BNC & kNN & LDA & BNC & kNN & LDA & BNC & kNN & LDA\\
\hline
  1 & 0.78 &\cellcolor{Gray} 0.97 & 0.39 & 0.75 & \cellcolor{Gray}0.93 & 0.48 & 0.77 & \cellcolor{Gray}0.98 & 0.32 & 0.78 &\cellcolor{Gray} 0.99 & 0.37\\
\hline
  2 & 0.78 & \cellcolor{Gray}0.90 & 0.55 & 0.80 &\cellcolor{Gray} 0.95 & 0.43 & 0.77 & \cellcolor{Gray}0.94 & 0.49 & 0.76 &\cellcolor{Gray} 0.94 & 0.49\\
\hline
  3 & 0.78 & \cellcolor{Gray}0.83 & 0.69 & 0.74 & \cellcolor{Gray}0.86 & 0.58 \cellcolor{Gray}& 0.83 & 0.82 & 0.65 & 0.74 &\cellcolor{Gray} 0.78 & 0.67\\
\hline
  4 & \cellcolor{Gray}0.82 & 0.69 & 0.56 & 0.70 &\cellcolor{Gray} 0.75 & 0.60 &\cellcolor{Gray} 0.76 & 0.69 & 0.61 &\cellcolor{Gray} 0.81 & 0.68 & 0.48\\
\hline
  5 & \cellcolor{Gray}0.75 &\cellcolor{Gray} 0.75 & 0.64 & \cellcolor{Gray}0.83 & 0.69 & 0.51 &\cellcolor{Gray} 0.83 & 0.61 & 0.54 & \cellcolor{Gray}0.85 & 0.66 & 0.57\\
\hline
  6 & \cellcolor{Gray}0.81 & 0.74 & 0.62 & \cellcolor{Gray}0.85 & 0.67 & 0.57 &\cellcolor{Gray} 0.83 & 0.69 & 0.60 &\cellcolor{Gray} 0.80 & 0.65 & 0.58\\
\hline
  7 & \cellcolor{Gray}0.73 & 0.67 & 0.66 & \cellcolor{Gray}0.78 & 0.63 & 0.63 & \cellcolor{Gray}0.79 & 0.59 & 0.65 &\cellcolor{Gray} 0.78 & 0.59 & 0.56\\
\hline
  8 & \cellcolor{Gray}0.76 & 0.70 & 0.72 & \cellcolor{Gray}0.87 & 0.67 & 0.65 & \cellcolor{Gray}0.85 & 0.67 & 0.78 & \cellcolor{Gray}0.69 & 0.65 &\cellcolor{Gray} 0.69\\
\hline
  9 &\cellcolor{Gray} 0.77 & 0.71 & 0.71 & \cellcolor{Gray}0.87 & 0.73 & 0.69 &\cellcolor{Gray} 0.87 & 0.69 & 0.74 & \cellcolor{Gray}0.73 & 0.67 & 0.71\\
\hline
  10 & \cellcolor{Gray}0.86 & 0.70 & 0.79 & \cellcolor{Gray}0.88 & 0.70 & 0.72 &\cellcolor{Gray} 0.93 & 0.72 & 0.72 &\cellcolor{Gray} 0.81 & 0.63 & 0.72\\
\hline
  11 & \cellcolor{Gray}0.84 & 0.69 & 0.76 & \cellcolor{Gray}0.86 & 0.69 & 0.71 &\cellcolor{Gray} 0.86 & 0.73 & 0.77 &\cellcolor{Gray} 0.85 & 0.67 & 0.73\\
\hline
  12 &\cellcolor{Gray} 0.84 & 0.69 & 0.78 & \cellcolor{Gray}0.86 & 0.77 & 0.74 &\cellcolor{Gray} 0.87 & 0.74 & 0.79 &\cellcolor{Gray} 0.81 & 0.68 & 0.74\\
\hline
  13 & \cellcolor{Gray}0.84 & 0.71 & 0.78 & \cellcolor{Gray}0.86 & 0.73 & 0.75 & \cellcolor{Gray}0.87 & 0.73 & 0.76 &\cellcolor{Gray} 0.84 & 0.72 & 0.69\\
\hline
  14 & \cellcolor{Gray}0.85 & 0.72 & 0.83 & \cellcolor{Gray}0.89 & 0.73 & 0.81 &\cellcolor{Gray} 0.82 & 0.76 & 0.75 & \cellcolor{Gray}0.83 & 0.76 & 0.80\\
\hline
  15 & \cellcolor{Gray}0.89 & 0.75 & 0.83 & \cellcolor{Gray}0.78 & 0.73 & 0.75 & 0.75 & 0.76 & \cellcolor{Gray}0.77 & \cellcolor{Gray}0.80 & 0.74 &\cellcolor{Gray} 0.80\\
\hline
  16 & \cellcolor{Gray}0.86 & 0.74 & 0.82 & \cellcolor{Gray}0.86 & 0.73 & 0.74 &\cellcolor{Gray} 0.82 & 0.76 & 0.75 &\cellcolor{Gray} 0.79 & 0.70 & 0.77\\
\hline
  17 &\cellcolor{Gray} 0.82 & 0.76 & 0.78 & \cellcolor{Gray}0.79 & 0.78 & 0.69 &\cellcolor{Gray} 0.78 & 0.75 & 0.71 &\cellcolor{Gray} 0.84 & 0.76 & 0.78\\
\hline
\end{tabular}
\end{table}

\newpage

\begin{figure}[!ht]
\begin{center}
\includegraphics[width=0.9\linewidth]{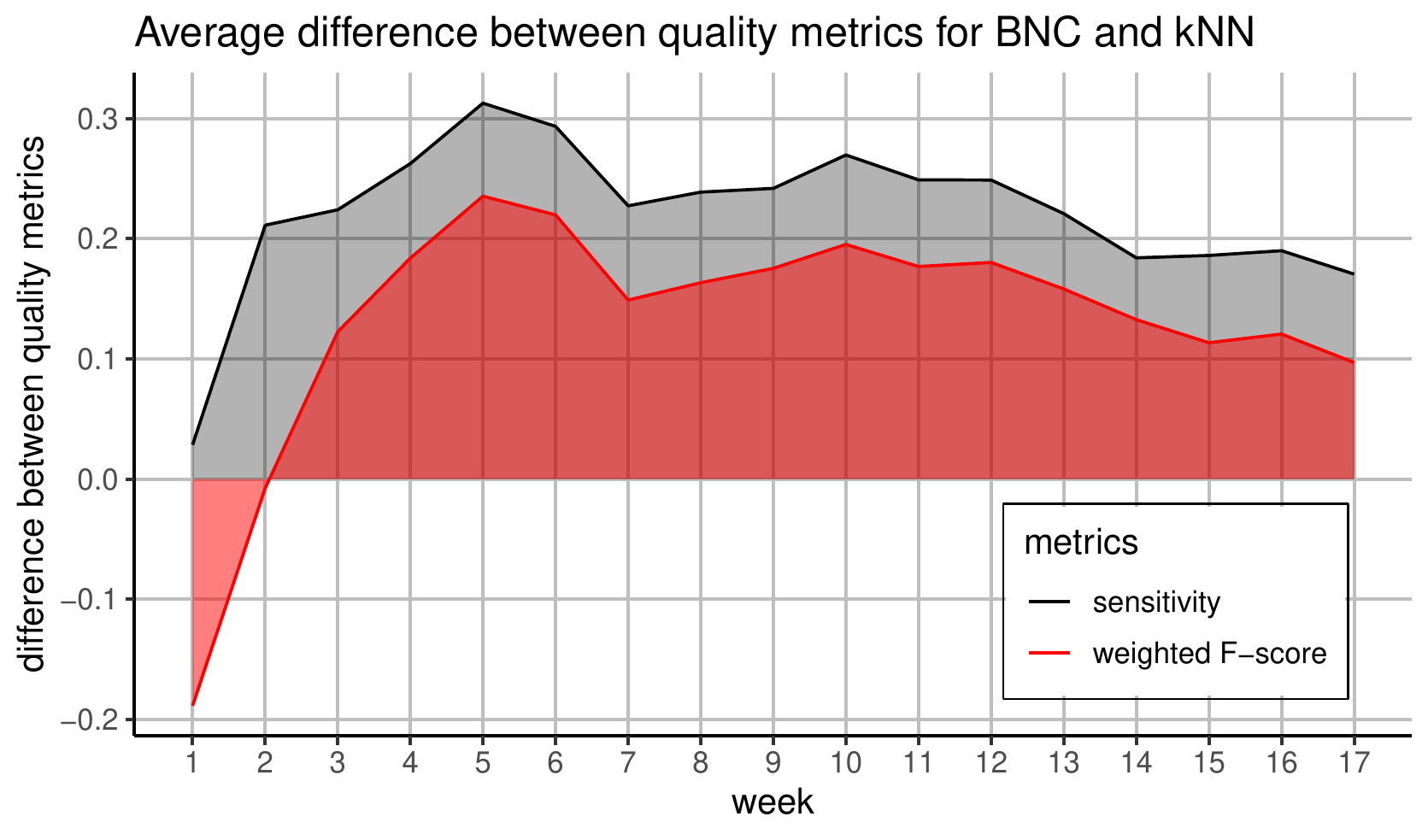}
\vspace{-0.5cm}

\caption{Average difference of classification performance metrics for BNC and kNN }
\label{Figure:excess-knn}
\end{center}
\end{figure}

\begin{figure}[!ht]
\begin{center}
\includegraphics[width=0.9\linewidth]{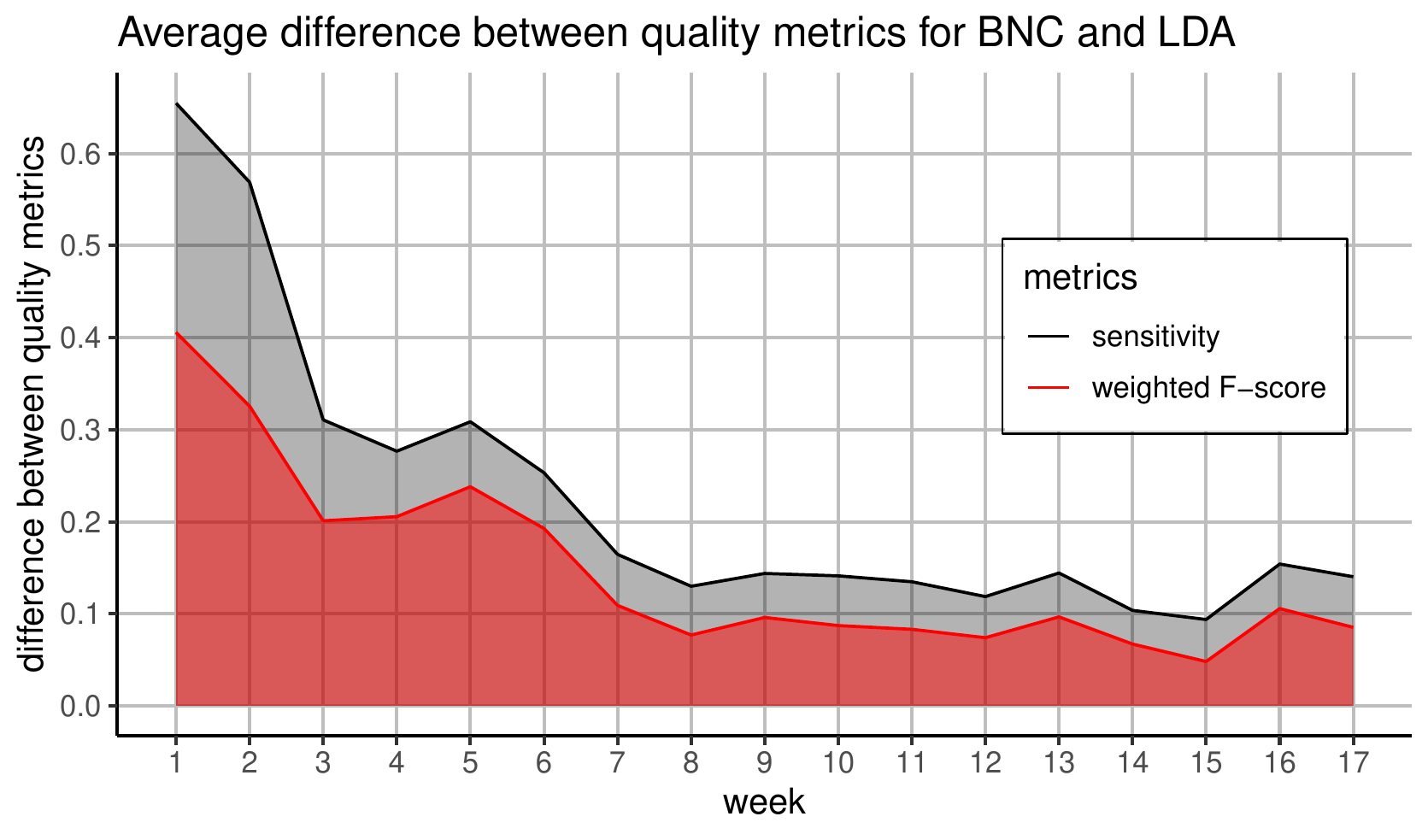}
\vspace{-0.5cm}

\caption{Average difference of classification performance metrics for BNC and LDA }
\label{Figure:excess-lda}
\end{center}
\end{figure}

The average values of sensitivity and weighted F-score over the fifteen testing sets were calculated for BNC, kNN and LDA algorithms. We compare the calculated values by plotting the excess (difference) of the values of the corresponding metrics of BNC model compared to kNN model (see Figure~\ref{Figure:excess-knn}) and that of BNC model compared to LDA model (see Figure~\ref{Figure:excess-lda}).


\vspace{-1cm}


\vspace{-0.5cm}


\newpage

\begin{figure}[ht]
\begin{center}
\includegraphics[width=0.95\linewidth]{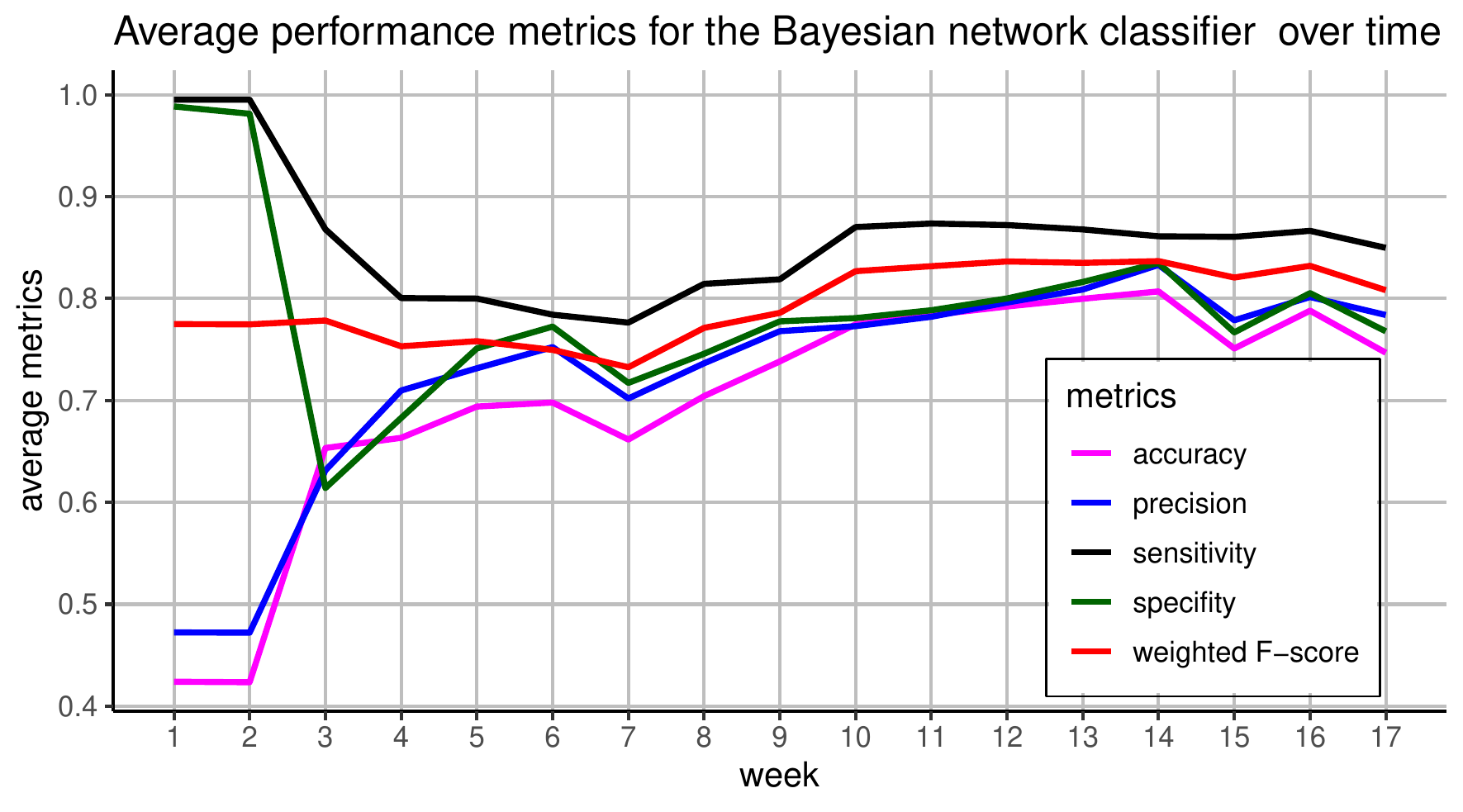}
\vspace{-0.5cm}

\caption{Average classification performance metrics for BNC model}
\label{Figure:metrics}
\end{center}
\end{figure}

Until the fourth week, most metrics for BNC model stay at an unsatisfactory level after what they start to increase and stabilize after week 9 (see Figure~\ref{Figure:metrics}) so that the quality of forecast could be assessed as appropriate (average sensitivity exceeds 85\% and average weighted F-score exceeds 80 \%).

\section{Conclusion}




The first aim of the present work was to develop an approach of student-at-risk detection via Bayesian networks which could serve as a basis for the development of a warning system of an academic course/module in a higher education institution. The general design of a Bayesian network and a classifier, based on this network is presented in Section~\ref{sec:gen_network}. 

The second aim was to demonstrate results of implementation of the approach on a certain course taught for the students of Siberian Federal University and to compare the quality of the developed Bayesian network classifier with the quality of other popular classifiers built on the same data and using the same features. For this purpose we have collected data on student performance and final grades for the course of Probability and Statistics (with the duration of one semester) for three consecutive years. On the formed dataset we have built three models (the designed Bayesian Network classifier, k-Nearest Neighbours classifier and Linear Discriminant Analysis classifier). The empirical study shows that from the point of view of accuracy, precision and specificity the  k-Nearest Neighbours Classifier with $k=3$ demonstrates the best result. However, the Bayesian network classifier provides a better quality of student-at-risk detection in regard to weighted F-score and sensitivity which both are more important  metrics for the considered problem of student-at-risk detection.

Starting from the tenth week, the constructed Bayesian network classifier is able to properly detect students, whose learning behavior is typical for unsuccessful (the average sensitivity of the classification performance exceeds 85\% and the average weighted F-score exceeds 80\% on the validation sets). This allows us to consider the approach to be perspective for the development of a warning system.

Yet, in the beginning of a semester the classifier tends to overestimate the risk of student failure (especially for the first 3 weeks, when the average precision is less than 70\%). We see two possible ways to solve this problem. First, we can possibly achieve better results by introducing into the model new predictors that might be valuable for our purposes, such as student grades for the prerequisite courses or those reflecting the history of student learning behavior. Second, we can possibly improve the quality of classification at early stages using ensemble methods. These are the objectives for future work.

\bibliography{Bayesian-network}

\end{document}